\documentclass[aps,prl,showpacs,letterpaper,twocolumn]{revtex4-1}
\usepackage{amsmath, amsthm, amssymb}
\usepackage{graphicx}
\usepackage[usenames]{color}

\newcommand{\rev}{\widetilde}
\newcommand{\work}{W}

\newcommand{\func}{{\mathcal A}}
\newcommand{\funcT}{{\mathcal G}}
\newcommand{\F}{\Lambda}				
\newcommand{\R}{{\rev{\Lambda}}}		
\newcommand{\RN}{{\rev{\Lambda^{\epsilon}}}}		
\newcommand{\Feq}{\lambda_{\rm{b}}}	
\newcommand{\Ieq}{\lambda_{\rm{a}} }	
\newcommand{\IeqF}{\Ieq,\F}	
\newcommand{\FeqR}{\Feq;\R} 
\newcommand{\FeqRN}{\Feq;\RN} 
\newcommand{\IeqFR}{\IeqF;\R}	
\newcommand{\IeqFRN}{\Ieq,\F^{\epsilon};\RN}
\newcommand{\Fx}{x(t_{\rm{b}})}	
\newcommand{\Neq}{\text{neq}}
\newcommand{\eq}{\text{eq}}

\newcommand{\Pneq}{P^{\Neq}}
\newcommand{\Peq}{P^{\eq}_{\lambda}}
\newcommand{\PFeq}{P_{\Feq}}
\newcommand{\PIeqF}{P_{\IeqF}}
\newcommand{\PIeqFN}{P_{\IeqF^{\epsilon}}}
\newcommand{\Pen}{P}
\newcommand{\dPen}{\delta P}

\newcommand{\ra}{\rangle}
\newcommand{\la}{\langle}
\newcommand{\kB}{k_{\text{B}}}
\newcommand{\kT}{\kB T}

\newcommand{\tI}{t_a}
\newcommand{\tF}{t_b}
\newcommand{\fric}{\zeta}

\definecolor{MyPurple}{rgb}{0.4,0.,0.35}
\definecolor{MyOrange}{rgb}{1.0,0.5,0.}

\def\be{\begin{equation}}
\def\ee{\end{equation}}

\def\md{\mathrm{d}}
\def\f{\frac}
\def\l{\left}
\def\r{\right}
\def\bla{\big\la}
\def\bra{\big\ra}

\begin{document}
\title{Near-equilibrium measurements of nonequilibrium free energy}
\date{\today}

\author{David A.\ Sivak and Gavin E.\ Crooks}
\affiliation{Physical Biosciences Division, Lawrence Berkeley National Laboratory, Berkeley, California 94720, USA}

\begin{abstract}
A central endeavor of thermodynamics is the measurement of free energy changes. Regrettably, although we can measure the free energy of a system in thermodynamic equilibrium, typically all we can say about the free energy of a nonequilibrium ensemble is that it is larger than that of the same system at equilibrium. Herein, we derive a formally exact expression for the probability distribution of a driven system, which involves path ensemble averages of the work over trajectories of the time-reversed system. From this we find a simple near-equilibrium approximation for the free energy in terms of an excess mean time-reversed work, which can be experimentally measured on real systems. With analysis and computer simulation, we demonstrate the accuracy of our approximations for several simple models. 
\end{abstract}
\preprint{}
\maketitle

Recent advances in nanotechnology make it increasingly possible to engineer  molecular scale structures for the deliberate and efficient manipulation of energy, matter and information on the nanometer scale. Artificial microscopic machines include heat pumps designed for very localized cooling; osmotic membranes built from carbon nanotubes; quantum logic gates designed to manipulate and stabilize quantum information; nanostructured thermoelectrics; devices for the capture and separation of carbon dioxide; and efficient photovoltaic solar cells.

Notably, molecular scale machines typically operate far from thermodynamic equilibrium, limiting the applicability of equilibrium statistical mechanics. Formulating a physically meaningful measure of the distance from equilibrium is itself an area of active research. Previous work developed a quantitative measure of the time asymmetry of ensembles of trajectories~\cite{Feng:2008ti}; in this Letter we quantitate the distance from equilibrium at one instantaneous snapshot, as expressed by a nonequilibrium generalization of free energy. While at equilibrium the free energy of a system is minimized (given the external constraints) and is often relatively easily measured, out of equilibrium no standard measurement technique exists, impeding the quantitative understanding of nonequilibrium behavior. To partially redress this deficit, we herein develop an experimentally tractable approach to measure the free energy of systems away from equilibrium. We find that to a strikingly good approximation, the difference between the free energy of a nonequilibrium ensemble and the equivalent system in equilibrium is determined by an excess mean time-reversed work, Eq.~\eqref{equ:FE}. The nonequilibrium probability of any given microstate is also well approximated by a similar excess mean time-reversed work, Eq.~\eqref{equ:pneqApprox}.

We consider a physical system in contact with a constant temperature heat bath with reciprocal temperature $\beta = (\kT)^{-1}$, where $\kB$ is Boltzmann's constant. The system has a collection of experimentally controllable parameters $\lambda$; for instance for a confined gas a control parameter could be the position of a piston dictating the volume of the chamber. To simplify the discussion throughout we refer to a single control parameter, though our analysis generalizes trivially to multiple control parameters. The free energy of the system, in or out of equilibrium, can be defined as~\cite{Gaveau:1997ul} $F \equiv \, \langle E \rangle - S/\beta$, for mean energy $\langle E \rangle \, \equiv \sum_x P(x) E(x)$ and entropy $S=-\sum_x P(x) \ln P(x)$ in natural units. Here, $x$ labels the microstates of the system. 

This generalizes the equilibrium free energy as a functional on the equilibrium distribution of microstates, to the nonequilibrium free energy as the \emph{same} functional on any (in general, nonequilibrium) distribution of microstates. Other rationales for calling this quantity a free energy are found in results for a system evolving according to a master equation. For such a system, when the control parameter is held fixed, this free energy difference is a nonincreasing function of time. If the system is allowed to fully equilibrate, it is equal to the total entropy produced (also known as the \emph{extropy})~\cite{Gaveau2002}. Hence this free energy difference has also been called an \emph{entropy deficiency}~\cite{Bernstein1972}. Equivalently, if the system is coupled to a mechanical system, this free energy difference equals the maximum work that can be done on that mechanical system while the original system relaxes to equilibrium (also known as the \emph{exergy})~\cite{Honerkamp02,Gaveau:2008hb}.

Interestingly, the free energy difference between two ensembles with identical values of the control parameter, one distributed among microstates according to the equilibrium probability distribution $\Peq(x) = \exp \{\beta\l[ F_{\lambda}^{\eq} - E_{\lambda}(x) \r]\}$ and one out of equilibrium and distributed according to $\Pneq$, is equal to the relative entropy  $D( \Pneq \| \Peq ) \equiv \sum_x \Pneq(x) \ln [\Pneq(x)/\Peq(x)]$ between the two probability distributions~\cite{Gaveau:1997ul}:
\begin{align}
D( \Pneq \| \Peq )
& = - S^{\Neq} -\sum_x \Pneq (x)\ \beta \l[F_{\lambda}^{\eq} - E_{\lambda}(x) \r] \nonumber \\
&= -S^{\Neq} - \beta F_{\lambda}^{\eq}+ \beta \la E_{\lambda} \ra_{\Neq} \nonumber \\
& =\beta\l(F_{\lambda}^{\Neq} - F_{\lambda}^{\eq}\r) \ .
\label{relative}
\end{align}
Here, $E_{\lambda}(x)$ is the energy of microstate $x$ given control parameter value $\lambda$, angular brackets with subscript ``neq'' denote an average over the nonequilibrium distribution $\Pneq$, $S^{\Neq}$ is the entropy of $\Pneq$, and $F_{\lambda}^{\eq}$ and $F_{\lambda}^{\Neq}$ are, respectively, the equilibrium and nonequilibrium free energies with control parameter value $\lambda$. Thus, in both a thermodynamic and information theoretic sense, this free energy difference between nonequilibrium and equilibrium ensembles measures a distance from equilibrium. 

A nonequilibrium ensemble is specified by a protocol~$\F$ that describes the history of the control parameter over some time interval: beginning at control parameter value $\Ieq$ at time $\tI$, the control parameter is changed according to $\F$ until it reaches value $\Feq$ at time $\tF$. In the corresponding time reversed protocol $\R$, the system starts at time $\tF$ with the final parameter $\Feq$ of the forward protocol, and then the controllable parameter retraces the same series of changes, in reverse, over a time interval of length $\tF-\tI$ to end at time $\tI$ with the initial value $\Ieq$ of the forward protocol. Measurements performed on a system using a pair of conjugate protocols $\F$ and $\R$ are related by~\cite{Crooks2000}, 
\begin{equation}
\bla \func \bra_{\lambda_a ;\F} = \bla \rev{\func} \, e^{-\beta \work} \bra_{\lambda_b ;\R} \Big/ \bla e^{-\beta \work }\bra_{\lambda_b ;\R} \ .
\label{equ:pea}
\end{equation}
Here, $\func$ is a measurement of the system (any real function of the phase space trajectory), $\rev{\func}$ is the corresponding time-reversed measurement (defined by $\func[{\bf x}] = \rev{\func}[\rev{\bf x}]$ where ${\bf x}$ and $\rev{\bf x}$ are a phase space trajectory and its time-reversal, respectively), and $\work$ is the work performed on the system during the (forward or time-reversed) protocol. The angled brackets indicate that measurements are averaged over an experimental protocol, specified by subscripts: the first subscript indicates the initial preparation of the system; the second subscript, after the semicolon, indicates the protocol during measurement. Thus ``$\Ieq;\F$'' specifies that the system is equilibrated with fixed parameter $\Ieq$ and then the properties of the system are measured while the system is driven with protocol $\F$, whereas ``$\Feq;\R$'' indicates initial equilibration at $\Feq$ followed by measurement during the time-reversed protocol~$\R$. If the preparation protocol is not explicitly stated, as is the case in many of our previous papers, then implicitly the system is prepared at equilibrium with the initial control parameter of the measurement protocol. 

We will use Eq.~(\ref{equ:pea}) to relate nonequilibrium probability distributions to moments of the work distribution. First, we replace the generic measurement $\func$ with a delta function $\delta\l[\Fx-x\r]$ of the final system microstate $\Fx$. This gives a relation between the nonequilibrium probability of a microstate, and a nonlinear average of the work performed on the system during the time-reversed protocol, starting from that microstate~\cite{Crooks2000}:
\begin{align}
\PIeqF(x) &= \bla \delta\l[\Fx-x\r] \bra_{\Ieq;\F} \nonumber \\
&= \bla \delta\l[\rev{x}(\tF) - x\r] e^{-\beta \work } \bra_{\Feq;\R} \Big/ \bla e^{-\beta \work }\bra_{\Feq;\R} \nonumber \\
&= \PFeq(x) \, \bla e^{-\beta\work} \bra_{x;\R} \Big/ \bla e^{-\beta \work }\bra_{\Feq;\R} \ .
\label{equ:pneq}
\end{align}
The subscript ``$x;\R$'' indicates initial preparation of the system in microstate $x$ and subsequent work measurement during protocol $\R$. Next, we rearrange the previous expression as in Ref.~\cite{Evans1995},
\begin{align}
\ln \frac{\PIeqF(x)}{ \PFeq(x)} & =\ln \frac{ \bla e^{-\beta\work} \bra_{x;\R} } { \bla e^{-\beta\work} \bra_{\Feq;\R}}\ , 
\label{equ:Pratio}
\end{align}
and factor out work averages to arrive at
\begin{align}
\ln \frac{\PIeqF(x)}{ \PFeq(x)} &= -\beta\l(\bla \work \bra_{x;\R} - \bla\work\bra_{\Feq;\R}\r) + \beta K_{x;\R}
\label{equ:Kintro} \\
\beta K_{x;\R} &\equiv \ln \frac{ \Big\la e^{-\beta\l(\work - \la\work\ra_{x;\R}\r)} \Big\ra_{x;\R} }{ \Big\la e^{-\beta\l(\work - \la\work\ra_{\Feq;\R}\r)} \Big\ra_{\Feq;\R}} \ .
\label{equ:Kdef}
\end{align}

Averaging over the nonequilibrium distribution gives the free energy difference
\be
F_{\Ieq,\F} - F_{\Feq} = -\l(\bla \work \bra_{\IeqFR} - \bla\work\bra_{\Feq;\R}\r) + \l\langle K_{x;\R} \r\rangle_{\F} \ .\label{equ:FEexact}
\ee
Here $F_{\Feq}$ is the equilibrium free energy under control parameter value $\Feq$, and $F_{\Ieq,\F}$ is the nonequilibrium free energy upon completion of protocol $\F$ following initial equilibration at $\Ieq$.  The subscript ``$\IeqFR$'' indicates initial preparation of the system by forward protocol $\F$ and subsequent work measurement during reverse protocol $\R$.

These relations for nonequilibrium probabilities and free energy are formally exact, yet impractical. In particular, the exponential averages in Eq.~(\ref{equ:Kdef}) are dominated by low dissipation realizations of the protocol, which are extremely rare~\cite{Jarzynski2006a}. 

To proceed further we develop a tractable approximation by examining, for a given nonequilibrium distribution $\PIeqF$ at the conclusion of protocol $\Lambda$, a family of nonequilibrium distributions $\PIeqFN(x) \equiv \PFeq(x) + \epsilon\l[\PIeqF(x)-\PFeq(x)\r]$. These distributions $\PIeqFN$ are produced by protocols $\Lambda^{\epsilon}$, which with probability $\epsilon$ reproduce the original protocol $\Lambda$ and with probability $1-\epsilon$ perform a reversible (quasistatic) protocol between the same two end points $\lambda_a$ and $\lambda_b$. In the near-equilibrium limit as $\epsilon \rightarrow 0$, expanding the relative entropy in $\epsilon$~\cite{Ingarden:1981uv} gives
\begin{align}
D\l(\PIeqFN \| \PFeq\r) &= \sum_x \PFeq(x)\l[1 + \dPen(x)\epsilon\r] \ \ln \l[1 + \dPen(x)\epsilon \r] \nonumber\\
&= \f{1}{2}\la\dPen^2\ra_{\Feq} \epsilon^2 -\f{1}{6}\la\dPen^3\ra_{\Feq} \epsilon^3 + O\l(\epsilon^4\r) \label{equ:rEExpand}
\end{align}
for the relative probability difference $\dPen(x) \equiv [\PIeqF(x) - \PFeq(x)]/\PFeq(x)$. The second line follows from Taylor expansion of the logarithm about $\dPen(x)\, \epsilon = 0$ and conservation of probability which imposes 
\be
\label{equ:probNorm}
\la \dPen \ra_{\Feq} = \sum_x \PFeq(x) \ \delta \Pen(x) = 0 \ .
\ee
Note that the leading-order term in Eq.~\eqref{equ:rEExpand} is one-half the Fisher information~\cite{Kullback:1951va}.

Under linear response~\cite{ZwanzigNESM}, deviations from equilibrium are expressed as an integrated response to external perturbation,
\begin{align}
&\l\la \Delta\funcT(\tF) \r\ra_{\Ieq;\F} = \beta \int_{t'=-\infty}^{\tF} \md t'\l[\lambda(t')-\lambda_b\r] \\
&\quad\quad\quad\quad\quad\quad\quad\quad\quad \times \f{\md}{\md t'} \l\la \delta\funcT(\tF) \, \delta B(t') \r\ra_{\lambda_b} \ .\notag 
\end{align}
Here, the control parameter $\lambda$ couples to the energy with conjugate force $B \equiv -\partial E/\partial\lambda$. $\l\la \Delta\funcT \r\ra_{\Ieq;\F}$ is the average deviation of measurement $\funcT$ (any real function of a point in phase space) at the conclusion of protocol $\F$ (running between times $\tI$ and $\tF$) from its equilibrium value at the final control parameter value $\Feq$. $\delta Y \equiv Y - \langle Y\rangle_{\lambda_b}$ is the instantaneous deviation of any variable $Y$ from its equilibrium value for control parameter value $\Feq$. $\l\la \delta\funcT(\tF)\, \delta B(t') \r\ra_{\lambda_b}$ is the covariance between the respective deviations from equilibrium averages of the measurement $\funcT$ and the conjugate force $B$, separated by time $\tF-t'$, at equilibrium under control parameter value $\Feq$. 

Integration by parts produces
\be
\l\la \Delta\funcT(\tF) \r\ra_{\Ieq;\F} = -\beta \int_{t'=-\infty}^{\tF} \md t' \f{\md\lambda(t')}{\md t'} \l\la \delta\funcT(\tF)\, \delta B(t') \r\ra_{\lambda_b} \ .
\ee
The boundary terms make no contribution because for an ergodic system all measurements separated by infinite time are uncorrelated, and thus $\lim_{t'\rightarrow-\infty} \l\la \delta\funcT(\tF) \delta B(t') \r\ra_{\lambda_b} = 0$.

Pulling the integral inside the average and substituting $\work_{\R} = \int_{\tI}^{\tF} \md t' \f{\md\lambda}{\md t'} B$ produces
\be
\l\la \Delta\funcT(\tF) \r\ra_{\Ieq;\F} = -\beta \la\funcT(\tF)\work_{\R} \ra_{\lambda_b} + \beta\la\funcT\ra_{\lambda_b}\la\work_{\R} \ra_{\lambda_b} \ .
\ee
Substituting $\funcT(\tF) = \delta[\Fx-x]$ gives
\be
\dPen(x) = -\beta \l( \l\la\work\r\ra_{x;\R} - \l\la\work\r\ra_{\FeqR} \r) \label{equ:lrW} \ .
\ee

This relation can also be derived by multiplying and dividing Eq.~\eqref{equ:pea} by $e^{-\beta\Delta F^{\rm eq}}$ and substituting the Jarzynski equality, $\l\langle e^{-\beta(\work-\Delta F^{\rm eq})} \r\rangle_{\FeqR}=1$, producing an alternative formulation of the path-weighted average,
\be
\langle \func \rangle_{\IeqF} = \l\langle \tilde{\func} e^{-\beta(\work-\Delta F^{\rm eq})}\r\rangle_{\FeqR} \ .
\ee
We subtract the final equilibrium average of $\func$ from both sides and substitute the Jarzynski equality again to get
\begin{align}
\langle \func \rangle_{\IeqF}& - \langle \func \rangle_{\Feq} = \\
&\l\langle \tilde{\func} e^{-\beta(\work-\Delta F^{\rm eq})}\r\rangle_{\FeqR} - \langle\func\rangle_{\Feq} \l\langle e^{-\beta(\work-\Delta F^{\rm eq})} \r\rangle_{\FeqR} \ .\notag
\end{align}
Substituting $\func = \delta[\Fx-x]$ produces
\be
\delta P(x) = \l\langle e^{-\beta(\work-\Delta F^{\rm eq})}\r\rangle_{x;\R} - \l\langle e^{-\beta(\work-\Delta F^{\rm eq})} \r\rangle_{\FeqR} \ .
\ee
Expanding near equilibrium to first order in $\work-\Delta F^{\rm eq}$, we arrive at~\eqref{equ:lrW}.

If instead of $\Lambda$ we apply protocol $\Lambda^{\epsilon}$, a similar derivation produces
\be
\dPen(x)\ \epsilon = -\beta \l( \l\la\work\r\ra_{x;\RN} - \l\la\work\r\ra_{\FeqRN} \r) \label{equ:lrWE} \ .
\ee
Averaging over the nonequilibrium distribution $\PIeqFN(x) = \PFeq(x)\l[1 + \dPen(x)\, \epsilon\r]$ gives
\be
\l\la\dPen^2\r\ra_{\Feq} \epsilon^2 = -\beta\l(\l\la\work\r\ra_{\IeqFRN}-\l\la\work\r\ra_{\FeqRN} \r) \ ,
\ee
where the $O(\epsilon)$ term on the left-hand side is zero by Eq.~\eqref{equ:probNorm}.

Substituting into the relative entropy expansion [Eq.~(\ref{equ:rEExpand})] and making use of the relation between relative entropy and free energy [Eq.~(\ref{relative})], the difference between the free energy of a nonequilibrium ensemble and the equivalent system at equilibrium is, to lowest order in~$\epsilon$, equal to minus one-half an excess mean time-reversed work:
\be
F_{\Ieq,\F} - F_{\Feq} \approx - \frac{1}{2} \Big( \la \work \ra_{\IeqFR}  - \la \work \ra_{\Feq;\R}\Big) \ .
\label{equ:FE} 
\ee
Here, finally, is our desired result. This free energy difference is readily measurable since it is minus one-half the average work $\la \work \ra_{\IeqFR}$ when the system is prepared with protocol $\F$ starting from equilibrium at $\Ieq$ and then driven with the time-reversed protocol $\R$, less the average work $\la \work \ra_{\Feq;\R}$ when prepared in thermal equilibrium at $\Feq$ and then driven with $\R$.

Comparing Eqs.~(\ref{equ:FEexact}) and (\ref{equ:FE}), our central result requires
\be
\l\la K_{x;\R}\r\ra_{\IeqF} \approx \f{1}{2}\l( \l\la\work\r\ra_{\IeqFR} - \l\la\work\r\ra_{\FeqR} \r) \ .
\ee
This is trivially satisfied when $K_{x;\R}$ is independent of $x$. Substituting this \emph{ansatz} into Eq.~(\ref{equ:Kintro}) gives a more manageable expression for the near-equilibrium probability distribution,
\begin{align}
\ln \frac{\PIeqF(x) }{ \PFeq(x)} &\approx -\beta \l[ \la \work \ra_{x;\R} - \f{1}{2}\l( \la \work \ra_{\FeqR} + \la \work \ra_{\IeqFR} \r) \r] \ .
\label{equ:pneqApprox}
\end{align}

Our derivation invokes the near-equilibrium limit, yet our expressions hold in wider contexts. Consider a system where they are exact: a micron-sized bead is suspended in water by an initially stationary optical laser trap with spring constant $k$, that is then translated at a constant velocity $v$, dragging the bead through the fluid with friction coefficient $\fric$. This system has been studied experimentally~\cite{Wang2002,Trepagnier2004} and can be modeled by a single particle undergoing diffusive Langevin dynamics on a moving, one-dimensional harmonic potential. The pertinent properties of the model have been analyzed~\cite{Mazonka1999,Horowitz2009}. Work distributions for a given initial particle position are Gaussian, with a mean work that depends linearly on the initial position of the particle relative to the center of the trap, and a position-independent variance. The equilibrium probability distributions are Gaussian, and the nonequilibrium probability distributions are also Gaussian, with the same variance, but shifted to a different mean relative to the equilibrium distribution. Consequently, our expressions for near-equilibrium probabilities and free energies are exact for this model at any driving rate, hence arbitrarily far from equilibrium. The free energy difference takes the simple form $\fric^2 v^2 / (4k)$.

\begin{figure}[t] 
 \centering
 \includegraphics{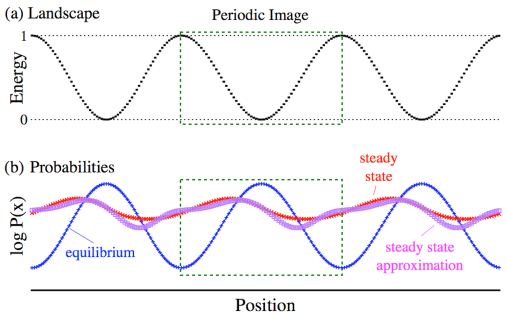}
 \caption{(Color online)  A simple driven system, amenable to numerical calculations. (a) Energy as a function of position. A single particle occupies a periodic, one-dimensional energy landscape. The position coordinate is discretized into $N_x$ uniformly spaced positions per period. Energy is also discretized, $E(x) = \lfloor N_e (1+ \sin (2\pi x/N_x))/2 \rfloor/N_e$, for position $x$ and number $N_e$ of discrete energy bins. $N_x$ and $N_e$ are increased until results do not change appreciably with a finer discretization. (b) The system is initially in equilibrium with an external heat bath~(\textcolor{blue}{$+$}). At each discrete time step, the particle attempts to move one step left, one step right, or remain in the same location with equal probabilities, and the move is accepted according to the Metropolis criterion~\cite{Metropolis1953}. Every $1/v$ time steps, the energy surface shifts one position to the right. To ensure fully time-reversible dynamics, we simulate $1/2v$ time steps, shift the potential, and simulate another $1/2v$ time steps before examining the nonequilibrium properties of the system. All figures are drawn in the rest frame of the potential. Eventually the spatial distribution across a single periodic image converges to a nonequilibrium steady state~(\textcolor{red}{$\times$}), approximated by Eq.~(\ref{equ:pneqApprox})~({\scriptsize \color{MyPurple}$\Box$}). Displayed results are for $v^*=24$, and reciprocal temperature $\beta = 4$ reported in inverse units of the energy difference between top and bottom of the potential.
}
\label{fig:systemSine}
\end{figure}

Empirically, our expressions are good approximations across a more general class of systems. To demonstrate this, we explore a system for which the steady-state probabilities, free energies, entropies and work distributions can be calculated exactly (within floating point accuracy). We simulate an overdamped particle diffusing over a periodic, sinusoidal, one-dimensional energy landscape. The particle begins at equilibrium for a fixed potential, and then the potential is translated at a constant velocity (Fig.~\ref{fig:systemSine}). 
Position and energy are discretized, hence all interesting properties of the system can be efficiently calculated using dynamic programming algorithms~\cite{Cormen2001, Crooks1999a}. See Fig.~\ref{fig:systemSine} for details. 

\begin{figure}[t] 
 \includegraphics{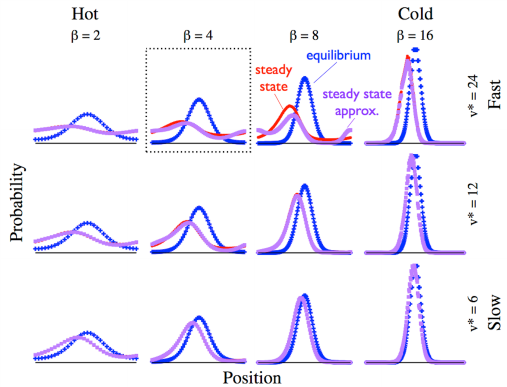}
 \caption{(Color online)  Equilibrium~(\textcolor{blue}{$+$}), steady-state~(\textcolor{red}{$\times$}) and approximate steady-state~({\scriptsize \color{MyPurple} $\Box$}) [Eq.~(\ref{equ:pneqApprox})] probability distributions, for the system described in Fig.~\ref{fig:systemSine}, at various driving rates and temperatures. Driving rates are reported in the dimensionless velocity $v^* \equiv v\ell/D$ for repeat length $\ell$ and diffusion coefficient $D$. The quality of our approximate distributions, including overall normalization, deteriorates at low temperature and high driving velocity. The dotted box highlights the conditions shown in Fig.~\ref{fig:systemSine}.}
\label{fig:3x4}
\end{figure}

\begin{figure}[t]
 \includegraphics{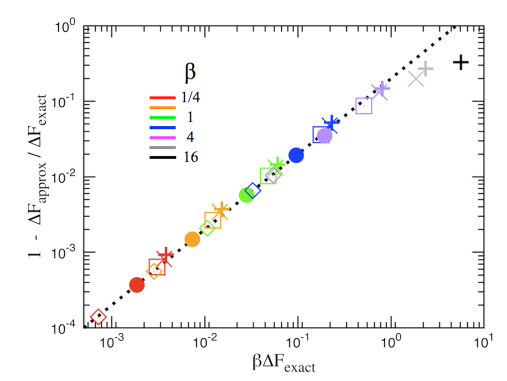}
 \caption{(Color online)  The approximate steady-state free energy difference per periodic image, $\beta\Delta F_{\textrm{approx}}$ [Eq.~(\ref{equ:FE})], is very close to the exact steady-state free energy difference $\beta\Delta F_{\textrm{exact}} = \beta\l(F_{\Ieq;\F} - F_{\Feq}\r)$, as shown by the fractional error $1 - \Delta F_{\textrm{approx}} / \Delta F_{\textrm{exact}}$ being much less than unity. Colors denote temperatures ranging from hot ($\beta = 1/4$, red, bottom left) to cold ($\beta = 32$, black, top right), with dimensionless velocity varying from $v^*=3$ (diamonds) to $v^*=48$ (crosses). Empirically for this system, $\beta\Delta F_{\textrm{approx}}$ is always less than $\beta\Delta F_{\textrm{exact}}$, and the fractional error shows a power-law dependence on exact free energy with exponent $\sim 1$ (dotted line plots $\, \tfrac{1}{5}\beta\Delta F_{\text{exact}}$). Note that before convergence to steady-state, fractional errors do not collapse onto a single curve even at low $\beta$.}
\label{fig:testAsymptote}
\end{figure}

Figs.~\ref{fig:3x4} and~\ref{fig:testAsymptote} demonstrate that for this model our steady-state probability [Eq.~(\ref{equ:pneqApprox})] and free energy [Eq.~(\ref{equ:FE})] approximations are accurate given slowly shifting landscapes or high temperatures, and only diverge significantly from the exact results in strongly driven systems. We also find that qualitatively similar results arise for aperiodic potentials and for different potential surfaces (data not shown). Fractional errors in the free energy estimate empirically equal $\sim0.2\ \beta\Delta F_{\text{exact}}$ in the near-equilibrium limit (Fig.~\ref{fig:testAsymptote}), suggesting the next term in a near-equilibrium expansion.

In this Letter, we have developed a practical method for measuring free energies in the near-equilibrium regime, and our simulation results indicate that the approximate relation between free energy and excess mean time-reversed work is accurate a substantial distance from equilibrium. Our analysis should be directly applicable to existing single-molecule experiments where the reverse protocol follows rapidly on the forward protocol, precluding equilibration~\cite{ClausenSchaumann:2000it}. We have concentrated on systems driven from equilibrium by a mechanical perturbation, but our relations could be generalized to other situations, for example a system driven by a temperature gradient~\cite{Jarzynski2004a}. Verifying our approximations in more complex systems will require independent measurements of nonequilibrium free energies; one possible approach for simple fluids would be to computationally estimate entropies from multiparticle distribution functions~\cite{Baranyai1989}. 

D.A.S. was funded by a National Science Foundation Graduate Research Fellowship. D.A.S. and G.E.C. were funded by the Office of Basic Energy Sciences of the U.~S. Department of Energy under Contract No. DE-AC02-05CH11231.

%

\end{document}